\documentclass[preprint2]{aastex}
\usepackage{hyperref,xcolor,natbib}
\hypersetup{
    pdfnewwindow=true,      % links in new window
    colorlinks=true,        % false: boxed links; true: colored links
    linkcolor=blue,         % color of internal links
    citecolor=black,        % color of links to bibliography
    filecolor=blue,         % color of file links
    urlcolor=blue           % color of external links
}

\def\be{\begin{equation}}
\def\ee{\end{equation}}

\providecommand{\aj}[0]{Astron. J.}
\providecommand{\apj}[0]{Astrophys. J.}
\providecommand{\apjl}[0]{Astrophys. J. Lett.}
\providecommand{\apjs}[0]{Astrophys. J. Supp. Ser. }

\providecommand{\aap}[0]{Astron. Astrophys. }

\providecommand{\mnras}[0]{Mon. Not. Roy. Astron. Soc. }

\providecommand{\nar}[0]{New Astron. Rev.}

\providecommand{\pasp}[0]{Publ. Astr. Soc. P.}

\begin{document}

\title{Galaxy Survey On The Fly: Prospects of Rapid Galaxy Cataloging to Aid the Electromagnetic Follow-up of Gravitational-wave Observations}

\author{I. Bartos$^1$, A.P.S. Crotts$^2$ and S. M\'arka$^3$}
\affil{$^1$Department of Physics, Columbia University, New York, NY 10027, USA}
\affil{$^2$Department of Astronomy, Columbia University, New York, NY 10027, USA}
\email{ibartos@phys.columbia.edu}

\begin{abstract}
Galaxy catalogs are essential for efficient searches of the electromagnetic counterparts of extragalactic gravitational-wave (GW) signals with highly uncertain localization. We show that one can efficiently catalog galaxies within a short period of time with 1-2 meter-class telescopes such as the Palomar Transient Factory (PTF) or MDM, in response to an observed GW signal from a compact binary coalescence. We find that a rapid galaxy survey is feasible on the relevant time scale of $\lesssim 1$\,week, maximum source distance of $>200$\,Mpc and sky area of 100\,deg$^2$. With PTF-like telescopes, even 1\,day is sufficient for such a survey. This catalog can then be provided to other telescopes to aid electromagnetic follow-up observations to find kilonovae from binary coalescences, as well as other sources. We consider H$\alpha$ observations, which track the star formation rate and are therefore correlated with the rate of compact binary mergers. H$\alpha$ surveys are also able to filter out galaxies that are farther away than the maximum GW source distance. Rapid galaxy surveys that follow GW triggers could achieve $\sim90\%$ completeness with respect to star formation rate, which is currently unavailable. This will significantly reduce the required effort and enhance the immediate availability of catalogs compared to possible future all-sky surveys.
\end{abstract}

\keywords{gravitational waves; methods: observational}

\section{Introduction}

\begin{figure}
\begin{center}
\resizebox{0.48\textwidth}{!}{\includegraphics{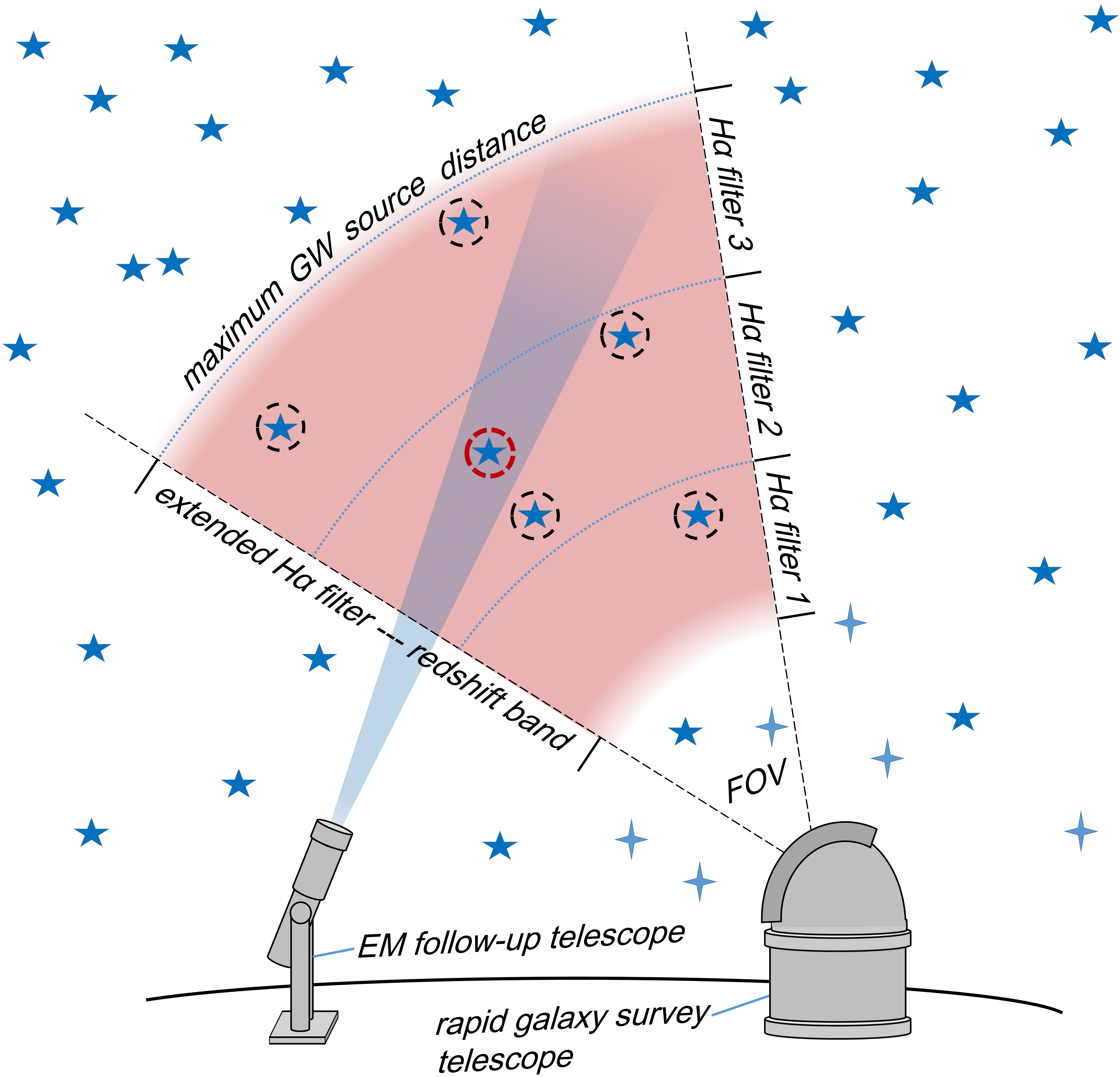}}
\end{center}
\caption{Schematic representation of a rapid galaxy survey and electromagnetic follow-up of a GW event. A telescope with large FOV (right) rapidly identifies galaxies within a distance range (red area) defined by the extended H$\alpha$ filter, or alternatively with multiple H$\alpha$ filters covering complementary redshift bands. The upper limit on the covered H$\alpha$-redshift ensures that bright galaxies beyond the GW sensitivity range do not contaminate the galaxy catalog, while the lower-redshift limit filters out local, e.g., galactic objects. A sensitive, narrow-FOV telescope then searches for the electromagnetic counterpart of the GW event in the directions of the identified galaxies.}
\label{JWSTmindetectionduration}
\end{figure}

The detection of gravitational waves (GW) from compact binary coalescences is expected to commence in the near future with the completion of the advanced LIGO detectors \citep{advancedLIGO0264-9381-27-8-084006} next year and advanced Virgo \citep{aVirgo} soon after \citep{2013arXiv1304.0670L}. GWs will enable the examination of binary coalescences, along with other phenomena, from new perspectives \citep{2009RPPh...72g6901A,2010CQGra..27q3001A,2013CQGra..30l3001B}.

To increase our confidence in the first detections, as well as to acquire complementary information from astrophysical sources, it will be critical to search for the electromagnetic or neutrino counterparts of GW signals. For the case of neutron star-neutron star and black hole-neutron star mergers, one of the most promising counterparts are kilonovae \citep{1998ApJ...507L..59L,2010MNRAS.406.2650M,2012ApJ...746...48M,2013ApJ...774...25K,2013Natur.500..547T}. Kilonovae are produced through radioactive decay of r-process elements, which are created in the neutron-rich material ejected during the merger. They produce quasi-isotropic emission that aids detectability compared to beamed, e.g., gamma-ray, emission. With $\gtrsim10^{41}$\,erg\,s$^{-1}$ peak luminosity in the near infrared and lasting for $\gtrsim$\,$1$\,week \citep{2013ApJ...774...25K}, kilonovae represent a bright and long emission that could be observed upon following up a GW signal candidate with a variety of telescopes.

The expected relatively poor localization of GW signals represents a significant difficulty for follow-up campaigns. Even with a network of 3+ GW detectors, typical events will be localized within $\sim$\,$10-100$\,deg$^2$. Early on, when only two detectors will be available, or for weak GW signals, the sky area will be $\sim$\,$100-1000$\,deg$^2$ (\citealt{2013arXiv1304.0670L}, and references therein). Such a large sky area (i) will result in a large number of false positive events \citep{2009aaxo.conf..312K}, and (ii) will make electromagnetic follow-up difficult with the sufficient depth.

Information on the location of galaxies within the localized sky area and within the sensitivity range of GW observations can significantly improve the potential for detection. Compact binary mergers are expected to occur in or near galaxies \citep{2010ApJ...708....9F}. Searching in the vicinity of galaxies is therefore sufficient to cover all mergers. In the following calculations we adopt a 200\,Mpc maximum range for GW detection from neutron star-neutron star mergers, while in Section \ref{section:distdependence} we then extrapolate the results to greater distances. 200\,Mpc is the fiducial direction- and orientation-averaged range for the advanced LIGO-Virgo detectors. Under favorable direction/orientation, neutron-star binary mergers will be detectable out to $\sim500$\,Mpc (e.g., \citealt{2013CQGra..30l3001B}). Within this range the expected, albeit uncertain, rate of neutron star-neutron star mergers is $0.4-400$\,yr$^{-1}$, most likely around $40$\,yr$^{-1}$ \citep{2010CQGra..27q3001A}.

To estimate the reduction of false positives due to using galaxy directions, one can consider the results of \cite{2013ApJ...767..124N}, who find that the number of galaxies within this volume is $\sim\,8$\,deg$^{-2}$ within 200\,Mpc. For 100\,deg$^2$ sky area, scaling the results of \cite{2013ApJ...767..124N} yields a total sky area occupied by the galaxies within 200\,Mpc to be $\sim$\,0.04\,deg$^2$, corresponding to $\mathcal{O}(10^3)$ reduction in the rate of false positives.

Similarly to follow-up surveys carried out for initial LIGO-Virgo \citep{loocUp,2014ApJS..211....7A}, galaxy locations can also help prioritize among directions followed up by telescopes with limited field of view (FOV). The advantage of prioritization will be most pronounced for these narrow-FOV telescopes. Consider, for example, a telescope with $\sim15' \times 15'$ FOV (Giant Magellan/IMACS, Long Camera Mode; \citealt{1998SPIE.3355..225B}). Such FOV corresponds to an average of $\sim0.5$ galaxies within $200$\,Mpc. Assuming uniform random galaxy distribution, the direction with the most galaxies within the FOV will have $\sim4$ galaxies, while $\sim60\%$ of the pointings will cover no galaxy. Prioritizing over which directions to follow up first can therefore significantly improve detection efficiency. Even for instruments with larger FOVs, prioritization will be important. For example, the BlackGEM Array \footnote{https://www.astro.ru.nl/wiki/research/blackgemarray} will consist of 60-cm telescopes, each with 2.7\,deg$^2$ FOV. Such FOV corresponds to $\sim20$ galaxies on average within $200$\,Mpc. For uniform galaxy distribution, the direction with the most galaxies within the FOV will have $\sim50\%$ more galaxies than a random direction, and about three times more galaxies than the direction with the least number of galaxies within the FOV.

Despite these potential advantages, galaxy catalogs are currently far from being complete for the relevant GW distance reach of $r\sim$\,$200$\,Mpc \citep{2011CQGra..28h5016W,2013ApJ...764..149M}, making the use of available galaxy catalogs less effective. For instance the galaxy catalog used for GW searches with initial GW detectors is estimated to be about $60\%$ complete with respect to B-band luminosity out to 100\,Mpc \citep{2011CQGra..28h5016W}.

In this paper we investigate whether a galaxy catalog can be assembled in 1\,week, or even 1\,day. We consider a hypothetical H$\alpha$ search with both the Palomar Transient Factory (PTF; \citealt{2009PASP..121.1395L}), and a 2-meter class telescope at the MDM observatory. We first limit the detection range to 200\,Mpc, and then generalize the search sensitivity to greater distances. Here, PTF and MDM are taken as examples to demonstrate the capability of existing and relatively easy-to-access instruments. Rapid surveys will likely be performed by a diverse group of telescopes. As this real-time survey will be critical for optimizing science return, the astronomical interest will be broad.

\section{H$\alpha$ survey requirements and completeness}
% describe what completeness is needed, why we use Malpha.

Star formation rate (SFR) is typically used in GW searches as the tracer of the rate of compact binary mergers \citep{2012A&A...541A.155A}. The merger rate, nevertheless, is expected to be somewhat delayed compared to SFR, and therefore may also be correlated to the total stellar mass in galaxies (e.g., \citealt{2010ApJ...725.1202L}).  In the following, we explore the prospects of an H$\alpha$ imaging survey, building on the close connection between H$\alpha$ emission and the ongoing SFR (e.g., \citealt{2003A&A...410...83H}). We adopt a survey depth of $F_{\rm lim,H\alpha}=10^{-15}$erg\,s$^{-1}$cm$^{-2}$ following \cite{2013ApJ...764..149M}, who find that $F_{\rm lim,H\alpha}$ corresponds to about $90\%$ completeness in SFR. This completeness threshold is chosen as it ensures the coverage of the majority of the sources, while it limits the observational cost, since the undetected $10\%$ H$\alpha$ luminosity is emitted by the faintest galaxies. Additionally, Metzger \emph{et al.} estimate that $F_{\rm lim,H\alpha}$ depth also renders a galaxy catalog about $50\%$ complete with respect to total stellar mass. We note here that, given available survey capability, it can be possible to go to even greater depths and and therefore further increase the completeness of the galaxy catalog.

\section{Sensitivity for one pointing}

We estimate the required duration of one pointing with a telescope to reach $F_{\rm lim,H\alpha}$ depth. We first consider a 2-meter class telescope, and for this example we adopt the parameters of the parameters of the Hiltner 2.4\,m telescope of the MDM Observatory \footnote{\url{http://mdm.kpno.noao.edu/index/MDM_Observatory.html}}. With such telescope size, $F_{\rm lim,H\alpha}$ corresponds to a photon detection rate of $\phi_{H\alpha}\sim 16$ ph/s. To measure the H$\alpha$ flux, we consider an extended H$\alpha$ filter with $[6530\,\mbox{\AA},6890\,\mbox{\AA}]$ that covers the H$\alpha$ lines within $[0\,$Mpc,$200$\,Mpc$]$. A similar idea of using multiple narrow-band H$\alpha$ filters for finding nearby galaxies was suggested earlier for PTF \citep{2013ApJ...767..124N}.

The background emissions we take into account are galaxy continuum emission and the night sky brightness. We use an R-band filter to estimate and subtract the background. To estimate the night sky brightness, we consider a sky area with $3$\,arcsec angular radius at 200\,Mpc, corresponding to a half-light radius of $\sim\,$3\,kpc, which is the typical galactic half-light radius in the R band \citep{2003MNRAS.341...33K}. Under favorable circumstances this corresponds to a photon rate of $\phi_{\rm sky}\sim 1.3\times10^3$ ph/s \citep{1998NewAR..42..503B}, which can increase by up to a factor of $\sim 7$ for unfavorable source direction, moonlight, and solar activity. To estimate galaxy brightness in the R band, we consider a galaxy at 200\,Mpc with R-band absolute magnitude $M_{\rm R}=-18$, corresponding to a photon rate of $\phi_{\rm galaxy}\sim 1.8\times10^3$ ph/s \citep{1998A&A...333..231B}. We choose this fainter $M_{\rm R}$ since fainter galaxies are more likely to be the ones undetected. For comparison, choosing $M_{\rm R}=-20$ would decrease the observable sky area below by a factor of two, i.e. the covered sky area is weakly dependent on our choice of $M_{\rm R}$. We note that fringing effects for this measurement are likely negligible (see, e.g., \citealt{2000SPIE.4008.1010C} for PTF and \citealt{doi:10.1117/12.2070014} for ZTF).

With these photon rates, we can estimate the minimum observation time $t_{\rm obs}$ required to reach $F_{\rm lim,H\alpha}$ depth with a signal-to-noise ratio SNR:
\begin{equation}
t_{\rm obs} \approx \mbox{SNR}^2\left(\frac{\phi_{\rm sky}+\phi_{\rm galaxy}}{\phi_{H\alpha}^2}\right)\left(\frac{\Delta\lambda_{H\alpha+}}{\Delta\lambda_{\rm R}}\right)^2,
\end{equation}
where $\Delta\lambda_{H\alpha+}=360$\,\mbox{\AA} and $\Delta\lambda_{\rm R}=1491$\,\mbox{\AA} are the widths of the extended H$\alpha$ and R-band filters, respectively. Requiring $\mbox{SNR}=5$, under favorable night sky conditions we find $t_{\rm obs}\approx40$\,s, while under ``typical" conditions $t_{\rm obs}\approx 80\,$s.

It is worth considering here the possibility that a kilonova is present in the pointing area that could in principle ``wash out" the galaxy as it increases the R-band background with little increase in H$\alpha$. To examine this scenario, we note that the R-band absolute magnitude $M_{\rm R}=-18$ considered for the faintest galaxies above corresponds to $\sim 6\times 10^{41}$\,erg\,s$^{-1}$ luminosity. This luminosity is greater than any of the kilonova peak luminosities considered in the literature \citep{2013ApJ...775...18B,2013Natur.500..547T,2014arXiv1411.3726K}. For comparison, peak R-band emission described recently by \cite{2014arXiv1411.3726K} is within $2\times10^{40}-4\times10^{41}$\,erg\,s$^{-1}$, or R-band absolute magnitude of $[-17.6,-14.3]$. This means that even at peak luminosity, kilonova emission will be significantly below the galaxy luminosity. An active kilonovae will therefore not affect the galaxy survey.

We now calculate the sensitivity using the parameters of PTF. We assume that the same filters are used in this case as well. With its 48-inch Samuel Oschin Telescope, the photon detection rate of PTF for $F_{\rm lim,H\alpha}$ flux is $\phi_{H\alpha}\sim 4$ ph/s. Using the same background photon rate as above, we arrive at $t_{\rm obs}\approx140$\,s under favorable conditions, and $t_{\rm obs}\approx300$\,s under typical conditions.

\section{Covered sky area for kilonovae}

To estimate the sky area that can be surveyed using MDM to find galaxies for kilonova searches, we consider $t_{\rm obs}=80$\,s for typical conditions from above, applied for both the H$\alpha$ and R-band observations. We further assume a 30-second CCD readout time, and 5-second slewing time between different directions. We consider a week-long as well as a day-long observation window after which kilonovae can still detectable, with 6 hours of observation per night. With the $24'\times24'$ FOV of MDM with its 8K CCD, we find the total surveyed sky area in 1\,week to be $\sim100$\,deg$^2$ (or $\sim15$\,deg$^2$ per day). We see that 1\,week is sufficient to find galaxies in the relevant sky area for GW observation, while even less time may be sufficient for better localized events, or if multiple MDM-like telescopes are available.

For PTF, we consider $t_{\rm obs}\approx300$\,s for both H$\alpha$ and R-band observations. We assume a readout time of 31\,s \citep{2009PASP..121.1395L}, and 5-second slewing. We take the Samuel Oschin Telescope's FOV on PTF being 8.1\,deg$^2$, we arrive at a covered sky area of $\sim1800$\,deg$^2$ over a one week period with 6-hours of observation per night (or $\sim260$\,deg$^2$ with 6\,hours of observations). We see that even 1\,day is sufficient for a PTF-like instrument to survey the sky area of interest following a GW signal. Additionally, this result indicates that a fraction of the 6-hour observation will be sufficient for a PTF-like telescope to cover the GW sky area, or a more detailed analysis can be performed in which more detailed information is obtained from the galaxies, such as their redshift.

\section{Distance dependence}
\label{section:distdependence}

While above we considered a fiducial distance of $r=200$\,Mpc, GWs are detectable under favorable directions and orientations for significantly greater distances. At design sensitivity, advanced LIGO-Virgo will be able to detect binary neutron star mergers out to $\sim450$\,Mpc \citep{2013CQGra..30l3001B}, and black hole-neutron star mergers even farther. It is therefore useful to examine the possibility of extending the distance reach of a rapid galaxy survey.

We estimated the sky area that can be cataloged by PTF or MDM as a function of $r$. The analysis was done similarly to the one presented above for 200\,Mpc. We assumed that $F_{\rm lim,H\alpha}$ scales with $r^{-2}$. We scaled the angular size of the background sky area $r^{-2}$, and modified the width of the H$\alpha$ filter to account for the redshift corresponding to $r$. We found that the covered sky area scales as $r^{-1.8}$. For a maximum source distance of $450\,$Mpc but otherwise using the same parameters as above, this corresponds to a covered sky area of $\sim400$\,deg$^2$ for PTF and $\sim\,30$\,deg$^2$ for MDM with 1\,week of observation (or $\sim60$\,deg$^2$ for PTF and $\sim\,3$\,deg$^2$ for MDM with 6 hours of observation). We see that even at this larger distance, $\sim$1-day will be sufficient for PTF-like telescopes for a rapid and comprehensive galaxy survey, while MDM will be able to build a catalog out to this larger distance for better localized GW signals.

\section{Conclusion}

With the sensitivity of present and planned telescopes, it will be difficult to follow-up a GW signal and efficiently scan $\gtrsim100$\,deg$^2$ of the sky for kilonovae. We investigated the capability of a 1 and a 2-meter-class telescope, PTF and MDM, to find galaxies over a large sky area of $\sim100$\,deg$^2$ within a limited time of $\lesssim1$\,week out to $r\geq200$\,Mpc. Such a rapidly assembled galaxy catalog could be used to guide electromagnetic follow-up searches of GW signals from compact binary mergers by significantly reducing the sky area that needs to be scanned. The 1-week time frame is aimed at aiding the observation of kilonovae, one of the most promising electromagnetic counterparts of compact binary mergers, expected to be bright for over a week. We adopted an H$\alpha$ survey that can be $\gtrsim90\%$ complete with respect to star formation rate out to the considered distances. We find that such a survey can find galaxies within a sky area of $\sim1800$\,deg$^2$ for PTF and $\sim100$\,deg$^2$ for MDM, within 1 week out to 200\,Mpc. We also find that even 1\,day of observation is sufficient to cover $\sim260$\,deg$^2$ for PTF out to 200\,Mpc, making PTF capable of assembling the required galaxy catalog in a matter of hours. Further, with such a sensitivity, rapid galaxy surveys with significantly greater source distances can be performed. For $450$\,Mpc, which is the maximum fiducial source distance for advanced LIGO-Virgo for neutron star-neutron star mergers, we find that PTF and MDM can survey $\sim400$\,deg$^2$ and $\sim\,30$\,deg$^2$, respectively, in 1\,week, and PTF will be able to survey $\sim60$\,deg$^2$ in 1\,day. This result shows that even at this largest fiducial distance, a PTF will be capable of assembling a galaxy catalog in $\sim1$\,day.

Many other telescopes, with suitable filters, would also be able to scan even larger sky areas over the allowed time window. For instance, PTF will soon be succeeded by the Zwicky Transient Facility (ZTF; \citealt{2014arXiv1410.8185B}, which will have a much larger FOV of $47$\,deg$^2$. With the appropriate filters it will be capable of cataloging galaxies within $\sim100$\,deg$^2$ in less than an hour. Other telescopes of interest include the Sloan Digital Sky Survey (SDSS; \citealt{1998AJ....116.3040G}), which could be mounted simultaneously with both the extended H$\alpha$ and R-band filters, allowing for synchronous observation with greater FOV than MDM. Further possibilities include, but are not limited to, CTIO 4m/DECam \footnote{\url{http://www.ctio.noao.edu/noao/content/Dark-Energy-Camera-DECam}}, Subaru/Suprime-Cam \footnote{\url{http://www.naoj.org/Introduction/instrument/SCam.html}}, SkyMapper \footnote{\url{http://rsaa.anu.edu.au/observatories/telescopes/skymapper-telescope}}, or, in the longer term, LSST \citep{2009arXiv0912.0201L}.

While the presented analysis demonstrates the feasibility of the rapid assembly of a galaxy catalog, there are several improvements that can be addressed in the future. For instance, the analysis above does not find out the distances of galaxies, beyond establishing that they are within 200\,Mpc. This can be remedied by, for instance, multiple, narrower H$\alpha$ filters that can be used to find galaxy distances. Further, the reconstructed distance of the GW source can also be used to tune the search. It will also be interesting to consider whether the rapid galaxy survey itself can be sensitive to also simultaneously detect the electromagnetic counterparts of GWs. To increase the completeness of the surveyed galaxy catalog with respect to stellar mass \citep{2013ApJ...764..149M}, one can further consider pursuing a similar rapid survey with wide FOV radio telescopes, such as ASKAP \citep{2008ExA....22..151J}. Nevertheless, one may not need very high completeness for effective follow-up observations \cite{2014ApJ...784....8H}. Further, the number of galactic foreground sources may be reduced by requiring a minimum redshift in designing the extended H$\alpha$ filter. Finally, rapid galaxy cataloging can be done with different techniques, and can involve multiple telescopes, further increasing the covered sky area or the completeness. This technique can aid GW electromagnetic follow-up observations even before comprehensive all sky surveys become available with suitable depths out to $\sim500$\,Mpc, and remedy the sensitivity limitation due to incompleteness of catalogs from the very first day of GW observations.

Rapid galaxy surveys will require (i) the initial investment of obtaining the appropriate extended or multiple H$\alpha$ filters, and (ii) a continued investment of several hours of telescope time per GW signal candidate of interest. The scientific return of propelling the first GW detections and the following chain of scientific discoveries will be well worth the effort. Further, an added beauty of real-time cataloging is that significant contribution can be realized by accessible telescopes, opening up the field for a broader range of collaborators worldwide.

% --- ACKNOWLEDGEMENT ------------------------------------------------------

The authors thank Marica Branchesi, Jules Halpern, Yiming Hu, Ilya Mandel, Zsuzsa Marka, Brian Metzger, Peter Shawhan and Laszlo Sturmann for their helpful comments. This article has been reviewed and approved for publication by the LIGO Scientific Collaboration. We are thankful for the generous support of Columbia University in the City of New York and the National Science Foundation under cooperative agreement PHY-0847182. The article has LIGO document number LIGO-P1400173.

\bibliographystyle{apj}

\end{document}